\documentclass[12pt]{article}
\usepackage[centertags]{amsmath}
\usepackage{graphicx}

\newcommand{\be}{\begin{equation}}
\newcommand{\ee}{\end{equation}}
\usepackage[centertags]{amsmath}
\usepackage{amsfonts}

\newcommand{\ba}{\begin{eqnarray}}
\newcommand{\ea}{\end{eqnarray}}

\begin{document}

\title{\bf  On $1+1$ Dimensional Galilean Supersymmetry in Ultracold Quantum Gases}
\author{ Gustavo~S.~Lozano$^a$\thanks{Associated with
CONICET},\, Olivier~Piguet$^b$\thanks{{}Supported
in part by the Conselho Nacional
de Desenvolvimento Cient\'{\i}fico e
Tecnol\'{o}gico CNPq -- Brazil.} ,\,
Fidel~A.~Schaposnik$^c$\thanks{Associated with CICBA}
\\ and
Lucas~Sourrouille$^a$
\\
{\normalsize \it $^a$Departamento de F\'\i sica, FCEyN, Universidad
de Buenos Aires}\\ {\normalsize\it Pab.1, Ciudad Universitaria,
1428, Ciudad de Buenos Aires, Argentina}
\\
{\normalsize\it $^b$Universidade Federal do  Esp\'\i rito Santo,
UFES, Vit\'oria, ES, } {\normalsize\it  Brasil}
\\
 {\normalsize\it $^c$Departamento de F\'\i
sica, Facultad de Ciencias Exactas}\\
{\normalsize\it Universidad Nacional de La Plata, C.C. 67, 1900 La
Plata, Argentina}
\\
{\footnotesize  lozano@df.uba.ar, opiguet@yahoo.com,
fidel@fisica.unlp.edu.ar, lsourrouille@yahoo.es} } \maketitle

\abstract{We discuss a $1+1$ dimensional  Galilean invariant model
recently introduced in connection  with ultracold quantum gases.
After showing its relation to a nonrelativistic $2+1$ Chern-Simons
matter system, we identify the  generators of the supersymmetry and
its relation with the existence of self-dual equations.}

\newpage


\vspace{1cm}

The study of supersymmetry started more than 30 years ago in the
context of {\em relativistic} field theories. They play today a
major role in the understanding of several properties of High Energy
Physics, ranging from phenomenological aspects of the Standard Model
of Electroweak Interactions to mathematical consistency of
Superstring theories~\cite{wess}.

From a mathematical point of view, supersymmetry is related to a
grading of the group of space-time symmetries of the theory and as
such, the concept of supersymmetry can be extended to other groups
beyond Poincar\'e.

The case where the space-time group is the Galilean Group is
particularly relevant as many condensed matter systems display
explicitly this symmetry.  The $3+1$ dimensional case was  first
considered in Ref.\cite{Puzalowski}. In Ref.\cite{leblanclozanomin},
the supersymmetric extension of the Galilean group in $2+1$
dimension was discussed and a field theoretical model possessing
this symmetry was explicitly built. The model is interesting as it
incorporates gauge fields. The dynamics of these gauge fields is
dictated by a Chern-Simons, which due to its topological character
 has a larger symmetry than the usual Maxwell action governing electromganetism.
It  is also possible to write for the bosonic sector of the model self-dual or Bogolnyi-Prassad-Sommerfeld (BPS) equations \cite{jackiwpi}, indicating that the relation between BPS equations and supersymmetry, characteristic of relativistic theories \cite{susy1},  also extends to the nonrelativistic domain. Galilean supersymmetry in $2+1$ was further studied in
\cite{2plus1}. Notice that in  the models considered in Refs.\cite{leblanclozanomin}, \cite{2plus1}, \cite{henkel}, the invariance group is indeed enlarged by the presence of
 dilatation and conformal symmetries leading to the study of Schroedinger super-algebras.

Another area where Galilean symmetry can be relevant is that
concerned with the study of perfect fluids. Also in this case
fermionic degrees of freedom cam be incorporated in the theory and
supersymetric models in $2+1$ and $1+1$ dimensions can be built
\cite{japo}, \cite{beja}, \cite{jarev}.

Very recently, a Galilean  model in $1+1$ dimension was considered
in Refs.\cite{vandoren1}, \cite{vandoren2} in the context of
ultracold boson-fermion mixture in one dimensional optical lattices.
This is indeed a very interesting proposal since according to these
authors it is possible to tune experimentally the parameters of the
model in such a way that the effective Hamiltonian describing the
dynamics of the vortex is indeed a $1+1$ dimensional Galilean
supersymmetric theory.

In this note we will study in more detail the structure of the
supersymmetry behind the model presented in \cite{vandoren1},
\cite{vandoren2}. In particular, we will find the  supersymmetry
transformations that leave the action invariant. A basic feature of
supersymmetric theories is that the {\it full} Hamiltonian can be
written as  an anticommutator of  supersymmetry generators.  We will
show that the charges generating these transformations in this
nonrelativistic model also satisty  this basic property (in
\cite{vandoren2} it was shown that the anticommutator of the charges
equals only the free  Hamiltonian).

In order to study the  supersymmetric structure of the theory, we
will make use of a relation of the $1+1$ model presented in
\cite{vandoren1},\cite{vandoren2} with the $2+1$ model studied in
\cite{leblanclozanomin}. This connection, which was already known
for the bosonic sector of the theory \cite{aglietti}, is not
necessary for establishing the main result of this note
(the correct  supersymmetry algebra), but in our  
opinion, the existence of  such a connection makes the model even more
interesting. On the other hand, it will lead us in a natural way to the
discussion of the relation between  supersymmetry and self-dual equations.

Let us then start by  considering the $2+1$ dimensional Chern-Simons
model coupled to nonrelativistic bosonic($\phi$) and fermionic
(down-spinor $\psi$)
 matter governed by the action,
\begin{eqnarray}
S&=& S_{cs}+\int d^3 x \left( i \phi^\dagger  D_t \phi + i
\psi^\dagger D_t \psi - \frac{1}{2m} ( D_i\phi )^\dagger  D_i\phi
-\frac{1}{2m}
( D_i\psi )^\dagger D_i\psi  \right.\nonumber \\
&-&  \frac{e}{2m} \psi^\dagger  F_{12} \psi  + \left. \lambda_{1}
(\phi^\dagger  \phi \phi^\dagger  \phi) + \lambda_{2} (\phi^\dagger
\phi  \psi^\dagger  \psi )\right) \nonumber \\
\end{eqnarray}
where the Chern-Simons  action is given by,
 \be S_{cs}=
 \frac{\kappa}{4} \int d^3x \epsilon^{\mu \nu \lambda}A_\mu F_{\nu
 \lambda}= \kappa \int d^3x \left( A_0 F_{12} + A_2 \partial_t A_1
\right) \ee

\begin{eqnarray}
F_{\mu \nu}=\partial_{\mu}A_{\nu}-
\partial_{\nu}A_{\mu}\,,
\;\;\;\;\;\
D_{\mu}= \partial_{\mu} + ieA_{\mu}
\end{eqnarray}
The metric tensor is  $g^{\mu \nu}=(1,-1,-1)$ and $\epsilon^{\mu\nu\lambda}$
is the totally antisymmetric tensor such that $\epsilon^{012}=1$.
We are choosing units such that $\hbar=1$.
We are including a  Pauli term for the fermion corresponding to  a
down-spinor.

The bosonic sector of the model ($\psi=0$) model was first considered in
Refs. \cite{hagen}, \cite{jackiwpi}. It is an explicit example of a
gauge theory  possessing  Galilean invariance and it provides a field theoretical formulation of the
 Aharonov Bohm problem.
Notice that there is no  
massless  particle associated to the vector field.  For the
particular relation of coupling constants \be
\lambda_1=\frac{e^2}{2m\kappa} \ee the model has self-dual equations
and many of the properties of its soliton solutions can be
established in detail \cite{jackiwpi}. The  fermionic generalization
of this model has been analyzed  in \cite{leblanclozanomin} where it
was shown that for a particular choice of coupling constants \be
\lambda_1=\frac{e^2}{2m\kappa}\,, \;\;\;\ \lambda_2=3\lambda_1 \ee
it exhibits an extended Galilean symmetry. Thus, the usual
connection between
 the existence of BPS equations and extended supersymmetry \cite{susy1} also holds in this nonrelativistic example.

In Ref.\cite{aglietti} a  reduction of the purely bosonic model to
$1+1$ dimensions was considered by assuming that the fields do not
depend on one of the spatial coordinates, say  $x_2$. One of the
main purposes in that
 work was to explore if the
peculiar  static properties of the parent $2+1$ dimensional model
were inherited by the $1+1$ reduction. The answer for this is
negative and, as it was discussed in detail in \cite{aglietti}, the
gauge fields can be eliminated from the theory without changing the
physical structure of the model.

Renaming   $A_2$ as $B$, and extending this procedure to the full model leads to an action that can be written as

\begin{eqnarray}
S&=& S_{rcs}+\int dxdt \lbrace  i \phi^\dagger D_t \phi + i
\psi^\dagger D_t \psi - \frac{1}{2m} ( D_x\phi )^\dagger D_x\phi
 - \frac{1}{2m} ( D_x\psi )^\dagger D_x\psi \nonumber \\ & &
-\frac{e}{2m}  \partial_x B \rho_f - \frac{e^2}{2m} B^2 \rho
+\lambda_{1} \rho_b^2 + \lambda_{2} \rho_b \rho_f\rbrace \label{action} \\
\end{eqnarray}
where $S_{rcs}$ is the reduced Chern-Simons action, or ``BF" term
\begin{eqnarray}
S_{rcs} = \kappa\int dxdt\left(  B\partial_tA_1 +  A_0 \partial_1 B\right)=\kappa \int dxdt B F_{01} \\
\end{eqnarray}
and we have introduced the matter densities,
\be
\rho_b=\phi^\dagger \phi \,,\,\,\,\;\;\;
\rho_f=\psi^\dagger \psi \,,\,\,\,\,\;\;\; \rho=\rho_b+\rho_f
\ee
Notice that the Gauss law constraint,
\be \partial_x B=\frac{e}{\kappa}  \rho \ee
can be solved as
\be B(x)=\frac{e}{2 \kappa} \int dz \epsilon(x-z) \rho(z)
\label{gauss} \ee
where $\epsilon(x)=\theta(x)-\theta(-x)$ is the odd step function

When $A_1$, $A_0$, and $B$ are set to zero the system has a
"trivial"  supersymmetry where bosons and fermions are interchanged
according to

\begin{eqnarray}
\delta_{1} \phi &=& \sqrt{2m} \eta_{1}^\dagger  \psi \;, \;\;\;
\;\;\; \delta_{1} \psi= - \sqrt{2m} \eta_{1} \phi\;,
\end{eqnarray}
 if $\lambda_2=2\lambda_1$. Here, $\eta_1$ is an infinitesimal
 Grassmann parameter.

This supersymmetry survives the incorporation of the additional
interactions if
\begin{eqnarray}
 \delta_{1} A_1 = \delta_{1}B&=& 0 \;, \;\;\; \;\;\;
\qquad\quad\ \delta_{1} A_{0}= \frac{e}{\sqrt{2m} \kappa} ( \eta_{1}
\phi \psi^\dagger  - \eta^\dagger _{1} \psi \phi^\dagger  ) \;
\label{var1}
\end{eqnarray}
and
\begin{eqnarray}
\frac{e^2}{2m\kappa} + 2\lambda_1 -\lambda_2 = 0\;.
\label{14}\end{eqnarray}
The model is also invariant under a second, less obvious,
 supersymmetry transformation given by

\ba
& &
\begin{array}{ll}
\delta_{2} \phi = \frac{i}{\sqrt{2m}} \eta_{2}^\dagger( D_x \psi -
eB\psi) \;   \qquad\qquad &\delta_{2} \psi = - \frac{i}{\sqrt{2m}}
\eta_{2} (D_x \phi+eB\phi)
\nonumber \\[3mm]
\delta_{2} A_{1} = -\frac{e}{\sqrt{2m} \kappa}\left( \eta_{2} \phi
\psi^\dagger - \eta_{2}^\dagger \psi \phi^\dagger \right) \; \;\;\;
&\delta_{2} B =  \frac{ie}{\sqrt{2m}\kappa}\left(\eta_{2} \phi
\psi^\dagger + \eta^\dagger _{2}
\psi \phi^\dagger \right)\; \nonumber \\
\end{array}
\\
& & \delta_{2} A_{0}   = \frac{ie}{( 2m )^{\frac{3}{2}}  \kappa} (
\eta_{2} \phi ( D_x \psi - eB\psi )^\dagger  + \eta^\dagger _{2} (
D_x \psi - eB\psi  ) \phi^\dagger  ) \;.
 \label{tiene}
\ea
provided that the coupling constants satisfy
 \be
\lambda_1=\frac{e^2}{2mk}
\label{16}\ee

 These transformations can be obtained via the dimensional reduction of the supersymmetry of the parent $2+1$ dimensional model \cite{leblanclozanomin} which in turn can be obtained via
the non-relativistic limit (contraction) of the associated
relativistic $2+1$ model \cite{relativistic}.

We will discuss next the relation of the model described by the action Eq.(\ref{action}) with the model
 discussed in Ref.\cite{vandoren2} where no  fields $A_0$, $A_1$ or $B$ appear.   The
 action of the model of Ref.\cite{vandoren2} without the chemical potential
 terms is,
\begin{eqnarray}
S& =& \int d^2 x \lbrace  i \phi^\dagger \partial_t \phi + i
\psi^\dagger \partial_t \psi - \frac{1}{2m} ( \partial_x\phi
)^\dagger
\partial_x\phi  - \frac{1}{2m} ( \partial_x\psi )^\dagger \partial_x\psi
+\lambda_1 \rho^2 \, .
\end{eqnarray}

It was shown in  \cite{aglietti}, \cite{oh}, for the bosonic sector of the  $1+1$ dimensional
model given by Eq.(\ref{action}), that the  field $A_1$ can be eliminated via a gauge transformation.
This property is also valid after the incorporation of fermions.
Indeed, after transforming the matter fields as
\be \phi(x)\rightarrow e^{-i\alpha(x)} \phi(x)\,, \,\,\,\,\,
\psi(x)\rightarrow e^{-i\alpha(x)} \psi(x) \ee with
\be
\alpha(x)=\frac{e}{2}\int dz \epsilon(x-z) A_1(z)
\label{alpha}\ee
and  using the  Gauss law, i.e, using the  explicit form of $B$ given
by equation (\ref{gauss}), the action can be written simply as

\begin{eqnarray}
S& =& \int dxdt \lbrace  i \phi^\dagger \partial_t \phi + i
\psi^\dagger \partial_t \psi - \frac{1}{2m} ( \partial_x\phi
)^\dagger
\partial_x\phi  - \frac{1}{2m} ( \partial_x\psi )^\dagger \partial_x\psi
-\frac{e^2}{2m \kappa} \rho \rho_f \nonumber \\
& &
+\lambda_{1} \rho_b^2
 + \lambda_{2}\rho_b\rho_f
- \frac{e^2}{2m} B^2 \rho
 \rbrace \nonumber \\
\end{eqnarray}
After elimination of the gauge field $A_1$ and  use of the
Gauss law, the SUSY transformations can be written as
\begin{equation}
\begin{array}{ll}
\delta_{2} \phi = \frac{i}{\sqrt{2m}} \eta_{2}^\dagger( \partial_x
\psi - eB\psi)  + i \delta_2 \alpha \; \phi\; ,  \qquad &\delta_{2} \psi = -
\frac{i}{\sqrt{2m}} \eta_{2} (\partial_x \phi+eB\phi) + i \delta_2 \alpha \; \psi\,,
 \label{tiene2}
\end{array}\end{equation}

The quantities $B$ and $\alpha$ should be considered  a functional of the matter fields, see
 Eq.~(\ref{gauss}) and Eq.~(\ref{alpha}).

Notice that the last term in the action is a constant of motion. Indeed
\ba
 & &\frac{e^2}{2m} \int dx_1 B^2(x_1) \rho(x_1)=
\frac{e^4}{8m\kappa^2} \int dx_1 dx_2 dx_3  \epsilon(x_1-x_2) \epsilon(x_1-x_3)
\rho(x_1) \rho(x_2) \rho(x_3) = \nonumber \\
& &
\frac{e^4}{24m \kappa^2} \int dx_1 dx_2 dx_3
\rho(x_1) \rho(x_2) \rho(x_3) =  \frac{e^4 N^3}{24m \kappa^2}
\ea
where  $N=\int dx \rho(x)$,  and use has been made of the identity

\be 
\epsilon(x_1-x_2) \epsilon(x_1-x_3) + \epsilon(x_2-x_3)
\epsilon(x_2-x_1) + \epsilon(x_3-x_1) \epsilon(x_3-x_2) = 1\,. \ee
Thus,  dropping this term, the action at the supersymmetric point (
given by Eqs. (\ref{14}) and (\ref{16})),
 can be written as,
\begin{eqnarray}
S& =& \int d^2 x \lbrace  i \phi^\dagger \partial_t \phi + i
\psi^\dagger \partial_t \psi - \frac{1}{2m} ( \partial_x\phi
)^\dagger
\partial_x\phi  - \frac{1}{2m} ( \partial_x\psi )^\dagger \partial_x\psi
+\lambda_1 \rho^2 \, .
\end{eqnarray}
 The model
written in this way is the one
 considered recently in \cite{vandoren2} in the context  of the dynamics
of vortices in boson-fermion mixtures.  In fact, this model was originally
considered by
Lai and Yang, \cite{lai}, (see also \cite{batch}).

In discussing the  supersymmetry algebra behind this model,
the following generators were given in Ref.\cite{vandoren2}, (up to normalization),

\be
 {Q}_1=-i \sqrt{2m}\int dx \psi^{\dagger} \phi \,\,\,\,\,\ R=-\frac{1}{\sqrt{2m}}\int dx \psi^\dagger \partial_x \phi
\ee The  $Q_1$ generator is related to the first of the
supersymmetries (\ref{var1}), and can be obtained via  the Noether
theorem. Poisson brackets can be defined as, \ba \lbrace F,G
\rbrace_{PB}=i\int dr \left( \frac{\delta F}{\delta \phi^\dagger(r)}
\frac{\delta G}{\delta \phi(r)}- \frac{\delta F}{\delta \phi(r)}
\frac{\delta G}{\delta \phi^\dagger(r)} - \frac{\delta^r F}{\delta
\psi^\dagger(r)} \frac{\delta^l G}{\delta \psi(r)}- \frac{\delta^r
F}{\delta \psi(r)} \frac{\delta^l G}{\delta \psi^\dagger(r)} \right)
\ea where the subscripts $r$ and $l$ refer to right and left
derivatives. In particular, \be
\lbrace\phi(x_1,t),\phi^*(x_2,t)\rbrace=-i \delta(x_1-x_2)\;\;\;\,
\lbrace\psi(x_1,t),\psi^*(x_2,t)\rbrace=-i \delta(x_1-x_2) \ee It is
easy to show that, \be \lbrace Q_1,Q_1^\dagger \rbrace= -2 i m \int
dx \rho \equiv -2 i M \,\,.\ee Nevertheless, the second generator
$R$ defined in \cite{vandoren2}, does not give
 transformations (\ref{tiene}). This $R$ charge is related to a symmetry
 of the free part of the theory (no interactions) and  its anticommutator gives
 only the free Hamiltonian, \be \lbrace
R,R^\dagger \rbrace=-i H_{free}=\frac{i}{2m} \int dx \;\;\;
(\phi^\dagger \partial_x^2 \phi + \psi^\dagger \partial_x^2 \psi)
\ee The  expression for the charge generating the second set of
transformations (\ref{tiene}) which correspond to a supersymmetry of
the full hamiltonian,  can be easily obtained by considering the
dimensional reduction of the corresponding charge in the $2+1$
model. This leads to \begin{eqnarray}
 {Q_2}&=&- \frac{1}{\sqrt{2m}}\int dx \psi^\dagger(\partial_x+eB)\phi
 \nonumber \\
  &=&- \frac{1}{\sqrt{2m}}
\int dx \psi^\dagger(x)(\partial_x+
\frac{e^2}{2 \kappa} \int dz \epsilon(x-z) \rho(z)
)\phi(x)
 \end{eqnarray}
Using the definition of  Poisson brackets, we can now calculate,
\ba \left\{Q_2,Q_2^\dagger \right\}&=&- \frac{i}{2m} \int  dx\left
 (\partial_x-eB) \psi^{\dagger} (\partial_x-eB) \psi + \right. \nonumber \\
 & & (\partial_x+eB) \phi^{\dagger} (\partial_x+eB)\phi
 -\frac{2e^2}{\kappa} \phi^\dagger \phi \psi^\dagger \psi  \; .
\label{last} \ea or
\begin{eqnarray}
\left\{Q_2,Q_2^\dagger \right\} &=&- \frac{i}{2m} \int
dx\left(\partial_x \psi^{\dagger} \partial_x \psi  + e^2 B^2 \rho_f
+ \partial_x \phi^{\dagger} \partial_x\phi + e^2B^2 \rho_b
\right. \nonumber\\
&& \left. -\frac{2e^2}{\kappa} \rho_f \rho_b - e
\partial_x B \rho_b +e \partial_x B\rho_f
\vphantom{\left(D_i
\psi\right)^{\dagger}}\vphantom{\frac{2e^2}{\kappa}} \right)
\;,\label{last2}
\end{eqnarray}
which, after using the Gauss law (eq.(\ref{gauss})) becomes \be
\left\{Q_2,Q_2^\dagger \right\}= -i H\;. \ee
Thus, the anticommutator of the charges is related to the {\em full} Hamiltonian.

Finally, the only remaining nonvanishing  bracket gives \ba
\left\{Q_1,Q_2^\dagger \right\} & =& -\frac{1}{2} \int d^2x
(\phi^\dagger (\partial_x-eB)\phi - ((\partial_x-eB)\phi)^\dagger
\phi + \psi^\dagger (\partial_x-eB)\psi -
((\partial_x-eB)\psi)^\dagger \psi) \nonumber \\
&=& -\frac{1}{2} \int d^2x
\phi^\dagger \partial_x\phi - \partial_x\phi^\dagger
\phi + \psi^\dagger\partial_x\psi -
\partial_x\psi^\dagger \psi)=-i P_1
\ea
being $P_1$ the momentum. We have used that,
\be
\int dx B(x) \rho=\frac{e}{2\kappa} \int dx dz \epsilon(x-z)\rho(x)\rho(z)=0
\ee
Then, with the use of the generators $Q_1$ and $Q_2$, the correct algebra is obtained,
\be
\lbrace Q_1,Q_1^\dagger \rbrace= -2 i m \int dx \rho= -2 i M\;\;\;
\left\{Q_2,Q_2^\dagger \right\}= -i H\;\;\;\;\left\{Q_1,Q_2^\dagger \right\}=-iP_1
\ee

In deriving the algebra, we have used canonical variables after the
implementation of Gauss law. In this way, the  SUSY charge $Q_2$ is a nonlocal function of the fields.
We could have arrived at the same algebra by keeping $A_1$ and $B$ as
conjugate variables,
\be
\lbrace A_1(x_1,t),B(x_2,t)\rbrace=\frac{1}{k} \delta(x_1-x_2)
\ee

To summarize, the connection of the Yang-Lai model to the $2+1$
dimensional Chern-{}Simons theory has helped us in identifying the
correct
 supersymmetry transformations and charges of the model. If we
consider the Yang-Lai model as a gauge ``BF" theory, then the SUSY
 transformations are local in space-time. If we instead
 eliminate the gauge fields
$A_0$ and $A_1$ together with the  scalar field $B$, then the Hamiltonian of the theory
becomes simpler, but the  supersymmetry charges become more complicated nonlocal functions of the physical fields.

The introduction of chemical potentials can be  achieved by
considering an effective potential \be \Omega=H-\mu_b N_b -\mu_f N_f
\,\,.\ee Then, as far as $\mu_b=\mu_f$, the transformations
generated by $Q_1$ and $Q_2$ are still formally a symmetry
$\{Q_i,\Omega\}=0$

As  discussed by \cite{oh}  the $B$ field plays also an important role
in the derivation
of the  self-dual equations. Indeed,  writing the action $S$ after
the implementation of the  Gauss law,

\begin{eqnarray}
 S& =& \int d^2 x \lbrace  i \phi^\dagger \partial_t \phi + i
\psi^\dagger \partial_t \psi - \frac{1}{2m} | (\partial_1+e\gamma_b
B)\phi|^2  - \frac{1}{2m} |( \partial_1+e\gamma_f B)\psi|^2
-\frac{e}{2m} \partial_1 B
 \rho_f + \nonumber \\
& &\frac{e}{2m}( -\gamma_b \partial_1 B \rho_b - \gamma_f
\partial_1 B \rho_f) +\lambda_{1} \rho_b^2 +
\lambda_{2}
\rho_b\rho_f\rbrace \nonumber \\
\end{eqnarray}
 leads to a Hamiltonian of the form,

\begin{eqnarray}
 H&=& \int d^2 x \lbrace    \frac{1}{2m} | (\partial_1+e\gamma_b
B)\phi|^2  + \frac{1}{2m} |( \partial_1+e\gamma_f B)\psi|^2
+\frac{e^2}{2m\kappa} \rho_T
 \rho_f \\
& &+\frac{e^2}{2m \kappa}( \gamma_b \rho_T \rho_b +\gamma_f \rho_T \rho_f)
-\lambda_{1} \rho_b^2 - \lambda_{2} \rho_b \rho_f \rbrace \nonumber \\
\end{eqnarray}
\begin{eqnarray}
 H&=& \int d^2 x \lbrace    \frac{1}{2m} | (\partial_1+e\gamma_b
B)\phi|^2  + \frac{1}{2m} |( \partial_1+e\gamma_f B)\psi|^2
\nonumber \\
& &+(\frac{e^2\gamma_b}{2m\kappa}   -\lambda_1)\rho^2_b +
(\frac{e^2}{2m \kappa}(1+\gamma_f+\gamma_b) -\lambda_2)\rho_b \rho_f
\rbrace \nonumber \\
\end{eqnarray}

Thus, for
\ba
\lambda_1 & = & \gamma_b \frac{e^2}{2m \kappa} \\
\lambda_2 & = &(1+\gamma_f+\gamma_b)\frac{e^2}{2m \kappa} \ea the
Hamiltonian is written as the sum of squares and minimum energy
configurations are such that they satisfy self-dual or BPS
equations, \ba & &(\partial_1+e\gamma_b
B)\phi=0 \\
& &
(\partial_1+e\gamma_f
B)\psi=0
\ea

The particular case $\gamma_b=\gamma_f=+1$ leads to the
supersymmetric case. Solitons solutions to this equations can be
found when $\lambda_1 \ge 0$, i.e, attractive self-interactions, and
they have been considered in detail in \cite{aglietti}, \cite{oh}

We have  thus discussed how a simple nonrelativistic supersymmetric
model in $1+1$ dimensions is related to a model of matter
interacting with gauge fields whose dynamics is  governed by a ``BF"
term. Although the gauge fields can be eliminated in terms of the
matter fields, their presence  helps in identifying
 the correct  supersymmetry transformations which in this case became a
{\it nonlocal} functional of the matter fields.

Due to the low  dimensionality, the model here  discussed might
provide the simplest theoretical realization of Galilean
Supersymmetry, and according to the authors of
Ref.\cite{vandoren1},\cite{vandoren2}, the model could be accessible
experimentally. From the theoretical point of view, it can  also
provide an interesting playground where to explore further the
relation between BPS states,
 supersymmetry and integrability.

\vspace{2cm}

G.S.Lozano thanks D.Jezek for interesting discussion on Bose
condensates and R.Jackiw for pointing out the existence of Galilean
supersymmetry in fluids. We specially thanks S.Vandoren for
interesting remarks on our manuscript. This work was partially
supported by CONICET grant PEI 6160.

\end{document}